\documentclass[a4]{elsart}

\usepackage{amssymb}

\usepackage{epsfig,rotating}
\usepackage{graphicx}

\usepackage{bm}
\def\NHN{{\textrm{N}-\textrm{H}\cdots\textrm{N}}}

\def\NN{\textrm{N}\cdots\textrm{N}}

\def\HN{\textrm{H}\cdots\textrm{N}}

\def\LNN{L(\textrm{N}\cdots\textrm{N})}
\def\cm{\textrm{cm}$^{-1}\,$}

\begin{document}

\begin{frontmatter}

\title{QM/MM Lineshape Simulation of the Hydrogen-bonded Uracil NH Stretching Vibration of the Adenine:Uracil Base Pair in CDCl$_3$}

\author[fu]{Yun-an Yan}
\author[fu]{,Gireesh M. Krishnan}
\author[ur]{,Oliver K\"uhn\corauthref{cor1}}
\address[fu]{Institut f\"{u}r Chemie und Biochemie, 
		Freie Universit\"{a}t Berlin, Takustra{\ss}e 3, 
		14195 Berlin, Germany}
\address[ur]{Institut f\"{u}r Physik, Universit\"{a}t Rostock, 
			Universit\"{a}tsplatz 3, 18055 Rostock, Germany}

\corauth[cor1]{Email: oliver.kuehn@uni-rostock.de }		

\begin{abstract}
A hybrid Car-Parrinello QM/MM molecular dynamics simulation has been carried out for the Watson-Crick base pair 
of 9-ethyl-8-phenyladenine and 1-cyclohexyluracil in deuterochloroform solution at room temperature.  The resulting
trajectory is analyzed putting emphasis on the N-H$\cdots$N Hydrogen bond geometry. Using an empirical correlation between the $\NN$-distance and the fundamental NH-stretching frequency, the time-dependence of this energy gap along the trajectory is obtained. From the gap-correlation function we determine the infrared absorption spectrum using lineshape theory in combination with a multimode oscillator model.  The obtained average transition frequency and the width of the spectrum is in reasonable agreement with recent experimental data. 
\end{abstract}

\end{frontmatter}

\section{Introduction}
The dynamics of the hydrogen bonds (HBs) linking complementary nucleic acid bases is an essential element responsible for the properties of DNA \cite{jeffrey97}. Among the various experimental methods used to study HBs, infrared (IR) spectroscopy  serves as an important means because of its remarkable sensitivity to HB formation and geometry \cite{taillandier92:307}. In particular, absorption lineshapes in the HB stretching region contain information on the strength of the HB, its anharmonic couplings to fingerprint modes via Fermi-type resonances and to low-frequency intra- and intermolecular HB modes. At the same time the widths of the absorption profiles give evidence for the interaction with the surrounding medium \cite{giese06:211}.
The details of HB lineshapes as well as the responsible dynamical processes can be investigated using nonlinear time-resolved IR spectroscopy \cite{nibbering07:619}. So far there are only a few ultrafast IR spectroscopic studies of DNA samples. Zanni et al. \cite{krummel06:13991} used two-dimensional IR spectroscopy in the carbonyl-stretching region to unravel coupling patterns sensitive to secondary structures  in guanine-cytosine bases. A related theoretical investigation based on a vibrational exciton model was given by Cho and coworkers (see, e.g., Ref. \cite{lee06:114510}). More recently, Heyne et al. employed the two-color pump-probe method to identify the symmetric NH$_2$ stretching fundamental transition in adenine-thymine oligomers \cite{heyne08:7909}. 

The complexity of DNA oligomers can be considerably reduced by focussing on the base pair building blocks. There are a number of  high-resolution IR+UV spectroscopic studies in the gas phase by the groups of  de Vries and Kleinermanns. In Ref. \cite{plutzer03:838}, for instance, it was shown that Watson-Crick (WC) pairing is not the most favorable case in the gas phase (see also simulation in Ref. \cite{krishnan07:132}). WC pairing can be enforced by substitutions as shown, e.g., for adenine and uracil derivatives in deuterochloroform solution  (9-ethyladenine, 1-cyclohexyluracil)  \cite{hamlin65:1734}. Recently, the first subpicosecond  time-resolved study of the vibrational relaxation of the H-bonded NH stretching fundamental transition in a substituted adenine-uracil pair has been reported by Wouterson and Cristalli \cite{woutersen04:5381}. In particular they observed that base-pairing decreases the relaxation time by a factor of three as compared to monomeric uracil. Since vibrational energy flow in H-bonded nucleic acid bases is of apparent importance for the understanding of DNA dynamics, not at least in the context of DNA photoprotection, there is a need to develop a microscopic picture for this process. 

In this contribution we will present the first step towards this goal which is the simulation of the NH stretching IR absorption lineshape around 3200 \cm. In doing so we will follow the argument of Baerends and co-workers, that  the electrostatic description of the HB in WC  base pairs is  questionable and charge transfer as well as resonance assistance occur which necessitates the quantum treatment of the electrons \cite{guerra99:3581}. For this purpose we employ Car-Parrinello Molecular Dynamics (CPMD)
\cite{car85:2471} which has been used previously to simulate IR spectra, e.g.,  of HBs formed between water molecules and uracil 
\cite{gaigeot03:10344}, of protonated water networks in bacteriorhodopsin \cite{rousseau04:4804}, of intramolecular H-bonds in a Mannich base in gas and crystalline phase \cite{jezierska07:205101,jezierska07:818,jezierska08:839} as well as of crystalline H$_5$O$_2^+$ClO$_4^-$ \cite{vener06:273} and picolinic acid N-oxide \cite{stare08:1576}.
In order to model the effect of the fluctuating deuterochloroform environment on the $\NHN$ HB dynamics we have chosen to use a QM/MM hybrid method \cite{laio02:6941} as detailed in Section \ref{sec:sim}. This will provide us with a classical trajectory from which time-dependent NH stretching frequencies are extracted using the correlation between the NH fundamental transition gap and the $\textrm{N}\cdots\textrm{N}$ distance \cite{novak74:177} (Section \ref{sec:res1}). In Section \ref{sec:res2} the frequency correlation function is determined and lineshape theory \cite{mukamel95} is used to obtain the IR absorption spectrum which is compared with the experiment results of Ref. \cite{woutersen04:5381}.
%
\section{Simulation Details}
\label{sec:sim}
Following the experimental setup of Ref.  \cite{woutersen04:5381} we will investigate the WC form of 
9-ethyl-8-phenyladenine (A) and 1-cyclohexyluracil (U)  (see, Fig. \ref{fig:structure}) solvated in CDCl$_3$. 
Our model is based on the hybrid QM/MM method as implemented in the Gromacs/CPMD interface \cite{biswas05:164114}.
All atoms of A:U pair other than those in the three substituents are dealt with by 
CPMD \cite{cpmd}. The substituents as well as the solvent are handled by Gromacs \cite{gmx31} using the OPLS all-atom force field \cite{jorgensen96:11225}. Since chemical bonds are cut by the QM/MM boundary, 
H atom capping is used to saturate the dangling bonds (see Fig.~\ref{fig:structure}).

In the simulation, one A:U pair is solvated in 100 $\textrm{CDCl}_3$ molecules within a box with dimensions 
$30.0~\textrm{\AA}\times23.5~\textrm{\AA}\times23.5~\textrm{\AA}$ (density is 0.1 M \cite{woutersen04:5381}). 
The QM part is placed in a $21.2~\textrm{\AA}\times15.9~\textrm{\AA}\times15.9~\textrm{\AA}$ box and the 
Becke exchange and Lee-Yang-Par correlation functional (BLYP) together with the plane wave basis 
set is used as implemented in the CPMD code. Further the Vanderbilt ultrasoft  pseudopotential 
is employed with a wave function cutoff of 30 Ry.  The MM molecular dynamics run is performed  at  298 K using a time step of 2 fs and the Car-Parrinello dynamics is integrated with a time step of 0.121 fs and a fictitious electron mass of 400 a.u.. 
The initial configuration corresponded to the geometry of the optimized gas phase base pair replacing solvent atoms in the equilibrated simulation box. Subsequently a 7.5~ps trajectory has been generated, the first picosecond of which was assigned for equilibration and not used for data analysis. 

It is well appreciated that there is a correlation between the HB length, i.e. here the $\NN$ distance, and the respective NH stretching frequency. Using the correlation curve obtained by Novak \cite{novak74:177} supplemented by more recent data we can establish the time-dependent fundamental NH stretching frequency, $\omega_{10}(t)$, along the trajectory and its correlation function
\begin{equation}
\label{eq:corr}
C(t) = \left\langle \left(\omega_{10}(t)-\langle\omega_{10}\rangle\right)
\left(\omega_{10}(0)-\langle\omega_{10}\rangle\right)\right\rangle \,. 
\end{equation}
Here, Ê$\langle\omega_{10}\rangle$ is the mean value along the trajectory. 
In passing we note that frequency-distance correlations also exist for O-H(D)$\ldots$O HBs \cite{mikenda86:1}. They have been used, e.g., in Ref. \cite{bratos04:197} to simulate lineshape functions for HOD in D$_{2}$O.

Adopting an adiabatic  two-level model and the Condon approximation where all other degrees of freedom except the NH stretching frequency are treated classically, the IR spectrum can be expressed via the lineshape function
\begin{equation}
\label{eq:gt}
g(t) \equiv \int^t_0d\tau\int^\tau_0d\tau^\prime\, C(\tau^\prime),
\end{equation}
as \cite{mukamel95}
\begin{equation}
\label{eq:irsp}
\sigma(\omega) = \frac{1}{\pi}\textrm{Re}\int^\infty_0 dt \,
	e^{i(\omega-\left\langle \omega_{10}\right\rangle) t - g(t)} \, .
\end{equation}
The correlation function, Eq. (\ref{eq:corr}), is a classical quantity and does not satisfy detailed balance. 
In principle detailed balance can be restored by applying certain correction factors. On the other hand, one might ask whether this matters at all for room temperature simulations which usually are characterized by rapidly decaying correlation functions. This problem has been addressed recently by Skinner and coworker for HOD in D$_{2}$O, who showed that quantum corrections have a modest influence on the IR lineshape only, but are important for nonlinear spectroscopies \cite{lawrence05:6720}.
Below we will  take into account quantum effects by adopting the multimode oscillator model \cite{mukamel95} for fitting the classical correlation function by the expression
\begin{eqnarray}
\label{eq:qcf} 
C(t) &= &C'(t)+iC''(t) \nonumber\\
&=&\sum_jS_j\omega^2_j\coth(\hbar\omega_j/2k_{\rm B}T)\cos(\omega_jt) + i \sum_jS_j\omega^2_j\sin(\omega_jt) \, .
\end{eqnarray}
Here, $S_j$ is the dimensionless Huang-Rhys factor giving the coupling strength between the two-level system and the mode with frequency $\omega_j$.  The real part $C'(t)$ is first used to fit the correlation function obtained from the molecular dynamics trajectory to a set of oscillators. Subsequently, $C''(t)$ is determined and the complex lineshape function is used to calculate the spectrum.
\section{Results and Discussions}
\subsection{Trajectory Analysis}
\label{sec:res1}
The geometry  is one of the most prominent characteristics of HBs and provides via correlation relations access to spectroscopic observables such as fundamental IR transitions \cite{novak74:177} or NMR chemical shifts \cite{pietrzak07:296}.
 In Fig. \ref{fig:traj}a we show the $\NN$ separation along the production part of the trajectory. Apparently, the HB length varies over a large range, that is, from 2.62~\AA ~to 3.94~\AA. The time average is 3.2~\AA ~for the $\NN$~ separation and 2.2 \AA ~for the $\textrm{H}\cdots\textrm{N}$ distance. The latter value is at the upper boundary of what has been reported for purine and pyrimidine crystals, that is,  $1.73 ~\textrm{\AA} < L(\HN) < 2.23 ~\textrm{\AA}$ with a mean of 
1.882~\AA{} \cite{jeffrey86:127}. 

Fig. \ref{fig:traj}b presents the distance between the H atom and the geometric center of the $\NN$ connecting line. The distance is defined to be negative if the H atom transfers to 
the adenine side. First, we notice that during most of the simulation interval the H atom is close to the uracil Nitrogen. Hydrogen transfer 
occurs in a short period of time only, that is, in the time interval $6.5~ \textrm{ps} < t < 6.56~ \textrm{ps}$. Apart from this short interval there is a clear correspondence  between $\NN$  and $\NHN$ distance changes. This indicates that the HB should be close to being linear. To clarify this point further the deviation of the H atom away from the linear motion in the HB defined as the difference between the sum  $L(\textrm{N}-\textrm{H}) + L(\textrm{N}\cdots \textrm{H})$ and the 
bond length $L(\textrm{N}\cdots \textrm{N})$, is drawn in panel (c) of Fig. \ref{fig:traj}.
Significant deviations occur  for times greater than $t=6.5~\textrm{ps}$ only and can be traced back to the H transfer event. Here, the 
H atom acquires additional kinetic energy after it transfers back to the uracil side. The reason lies in the fact that  the potential energy curve  is an asymmetric double well with the higher energy at the adenine side. The additional kinetic energy will lead to a substantial out of line motion until it is dissipated to the A:U pair and the solvent, as shown in the window around $t=7~\textrm{ps}$ of this panel. 
Except for the period of H transfer and relaxation thereafter 
the difference is smaller than 0.09 \AA~for 95\% of the simulation time and 
smaller than 0.14 \AA ~for 99\% of the time. The average deviation 
is 0.034 \AA~ for times before the H transfer and 0.044~\AA~
for the whole simulation. Compared with the average NH distance 
(1.05 \AA), the deviation is small and the HB can indeed be considered to be linear.

In a next step we will correlate the $\NN$ separation to the NH stretching frequency using the experimentally established correlation of Novak 
 \cite{novak74:177}  which we supplement by two additional data points for small distances (see caption Fig. \ref{fig:corr}). The frequency-distance correlation is shown in Fig. \ref{fig:corr}a together with a linear regression fit to the function
\begin{equation}
\label{eq:fit}
f(r) = \omega_\infty/2\pi c \, {\rm erf}\left(\sum_{i=0}^2 a_i r^i\right),
\end{equation}
where $\omega_\infty$ is the  free NH stretching frequency in gas phase  (for parameters see figure caption).
As a note in caution we emphasize that this type of correlation curve is of approximate nature. 
On the experimental side, the spectrum is often very broad and bears sub-band structure 
which makes the assignment difficult.  On the computational side, this model assumes that the H atom is 
attached the original site, therefore, there is no H transfer at all. 
Of course, this is not the case throughout our simulation and we will have to exclude the corresponding configurations. Further we note that Eq. (\ref{eq:fit}) also should not  be used for very strong HBs, since there the transition frequencies are very sensitive to the details of the particular potential energy surface as well as to zero point energy effects (see, e.g. Ref. \cite{asmis07:8691}).

With help of the frequency-distance correlation,  the transition frequency evolution along the trajectory can be obtained
from the  HB $\NN$ length. The result is shown in Fig. \ref{fig:corr}b which also displays the rescaled HB length to emphasize the correlation between stretching frequency and HB  length. Notice that the correlation curve levels off at large $\NN$ distances such that the frequency essentially stays constant. The average frequency along the trajectory in the interval 1.0-6.2 ps is 3227~\cm. 
\subsection{IR Lineshape}
\label{sec:res2}
Having calculated the time-dependent frequency $\omega_{10}(t)$ we can proceed to determine the correlation function, Eq. (\ref{eq:corr}).
This is done by assuming ergodicity, that is, $C(t)$ is obtained along the trajectory via 
\begin{equation}
\label{eq:ccal}
C(t) = \frac{1}{T_f - T_i - t} \int^{T_f - t}_{T_i} d\tau
	(\omega_{10}(t+\tau) -\langle\omega_{10}\rangle)
	(\omega_{10}(\tau) -\langle\omega_{10}\rangle),
\end{equation}
where $[T_i, T_f]$ is a certain propagation time interval. Concerning this time interval 
one has to be cautious to ensure that the time $T_i$ is chosen such that the system has equilibrated from the initial configuration.
The impact of different choices of the time interval is shown in Fig.~\ref{fig:spec}a and b. One finds that when $T_i$ varies from $T_i=1.0~\textrm{ps}$ to $T_i=1.4~\textrm{ps}$ the differences of the correlation function are minor for  $t<1.8~\textrm{ps}$ and quite small for $t>1.8~\textrm{ps}$. This finding indicates that as far as the simulation of $C(t)$ is concerned the system is indeed sufficiently equilibrated at $t=1.0~\textrm{ps}$. Panel (b) shows that for $T_i=1.0~\textrm{ps}$, the result for $T_f=6.0~\textrm{ps}$
is almost identical to that for $T_f=6.2~\textrm{ps}$ but quite different
from that obtained using  $T_f = 7.2~\textrm{ps}$. The reason lies in the previously mentioned fact that
the H transfer which occurs at about $t=6.6~\textrm{ps}$ 
leads the system away from the uracil NH equilibrium site. In other words, H transfer is a singular event in the considered time interval and in order to make any statement concerning its influence one should have at hand a much longer trajectory sampling many of such H transfer events. In addition the correlation in Fig. \ref{fig:corr}a has been developed for local NH stretching vibrations and becomes questionable if the N-H bond is replaced by a N$\cdots$H bond.
Combining the two panels, one can draw the conclusion that the correlation function after 
time $t=1.8~\textrm{ps}$ is sensitive to the time interval and
is not reliable. Based on this consideration, the time range 
$1.0~\textrm{ps} \leq t \leq 6.2~\textrm{ps}$ will be used for 
the time average. 

To proceed we will fit $C(t)$ using the multimode oscillator model, Eq. (\ref{eq:qcf}), as outlined in Section \ref{sec:sim}. The frequency interval is dictated by time-step and interval of the correlation function, i.e. 80-1770~\cm, discretized into 20 modes whose frequencies and coupling constants are fitting parameters. Given the number of parameters, the correlation function can be fit almost perfectly in the interval up to 1.8 ps.  The fitted coupling strength is a smooth function 
of the frequency, as shown in Fig.~\ref{fig:spec}c. Specifically we find that the coupling between the NH vibration of the A:U pair and all other degrees of freedom is dominated by low-frequency modes. This includes A:U modes modulating the HB length ($\NN$ distance). Preliminary normal mode calculations for the gas phase A:U pair show that the shoulder around 400 \cm could also be due to various modes having $\NN$ stretching character. In passing we note that this shoulder is reproducible for different sets of starting frequencies in the fitting.

Finally, in Fig. \ref{fig:spec}d we compare the   simulated and measured \cite{woutersen04:5381} IR absorption spectrum. The experimental spectrum peaks at 3185 \cm and has a full-width at half maximum (FWHM)  of 53 \cm. The average transition frequency of our simulation has been 3227 \cm which upon incorporating the harmonic correction factor shifts to 3209 \cm, i.e. the difference with respect to the experiment is only 1\%. 
The calculated FWHM is 39 \cm which is 24\% smaller than the experimental value. Both spectra show some asymmetry with respect to the high energy side. We note that without the quantum correction the linewidth reduces to 32 \cm. Besides the perhaps surprisingly good agreement between theory and experiment the analysis in Fig. \ref{fig:spec}d allows to assign the spectrum of modes coupled to the NH stretching vibration. One notices that this spectrum is cut-off at high frequencies. Therefore, contributions from the first overtone of the NH bending vibration are not explicitly included. It can be argued that possible Fermi resonances are implicitly included in the frequency-distance correlation. However, to establish the role of such a Fermi resonance a model which includes both vibrations, e.g., by calculating respective potential energy surfaces for snapshots along the trajectory, should be used. 

In summary we have presented a QM/MM approach to the calculation of the IR lineshape of the H-bonded NH stretching vibration of an A:U pair in CDCl$_3$ solution. The time-dependence of the transition frequency was obtained using the $\NN$-distance-frequency correlation from Novak \cite{novak74:177} supplemented by more recent data and fitted to a function which contained the free NH vibration as a limiting case. The obtained average $\NN$ distance of 3.2 \AA{} corresponds to a transition frequency of 3344 \cm, which is slightly larger than the experimental value (3185 \cm  \cite{woutersen04:5381}). However, using lineshape theory together with a multimode oscillator model an IR spectrum has been obtained which is quantitative as compared to the experiment within 1\% for the line position and 24\% for the FWHM. The spectral density responsible for the width of the spectrum is dominated  by modes in the range below 200 \cm. 
\section*{Acknowledgments}
We are indebted to Dr. H. Naundorf (Berlin) for his continuous support in the software installation and to Dr. S. Woutersen (Amsterdam) for the helpful discussion concerning the NH assignment. This work has been financially supported by the Deutsche Forschungsgemeinschaft (Sfb450).

%
%
\newpage
\begin{figure}[ht]
\caption{Structure of Watson-Crick pair formed by 9-ethyl-8-phenyladenine and 1-cyclohexyluracil and separation into QM and MM regions. The three H atoms bonded to $\textrm{N}_9$ and $\textrm{C}_8$ of adenine and $\textrm{N}_1$ of uracil,
respectively, are capping atoms in the QM/MM treatment.}
\label{fig:structure}
\end{figure}
\begin{figure}[ht]
\caption{Geometric parameters of the $\NHN$~ hydrogen bond along the trajectory:
(a) HB length; the horizontal line indicates the time average,
(b) distance between the H atom and the center of the $\NN$ connecting line,
(c) out-of-line motion of the H atom as measured by the difference
$L(\textrm{N}-\textrm{H}) + L(\textrm{N}\cdots \textrm{H}) - L(\textrm{N}\cdots \textrm{N})$.
Notice that the first picosecond of the trajectory has been assigned for equilibration from the initial configuration.
}
\label{fig:traj}
\end{figure}
\begin{figure}[ht!]
\caption{
(a) Empirical frequency-distance correlation function. The dots give the frequencies obtained from crystals containing
    intermolecular $\NHN$ HBs by Novak \cite{novak74:177} which are supplemented by the points for 4-aminopyridine hemiperchlorate (2240 \cm, 2.698 \AA{}) \cite{roziere80:6117} and phtalazine hemiperchlorate (2370 \cm, 2.656 \AA{}) \cite{grech80:495}. The solid line corresponds to the fitting function in Eq.~(\ref{eq:fit}) using the free NH stretching frequency of uracil in the gas phase ($\omega_\infty/2\pi c =$ 3436~\cm~\cite{colarusso97:39}). The fit parameters for Eq.~(\ref{eq:fit}) are $a_0 = 7.0911$, $a_1 =-5.7941~\rm{\AA}^{-1}$, and $a_2 = 1.2711~\rm{\AA}^{-2}$.
(b) Transition frequency $\omega_{10}$ as a function of time. 
    Solid line:  frequencies fitted from the frequency-stretching correlation function,
    Dashed line: HB length $\LNN$ shifted and rescaled such as to enable correlation with the frequency,
    Dotted line: 3216~\cm, the time average of the frequency along the trajectory.}
\label{fig:corr}
\end{figure}
\begin{figure}[ht!]
\caption{Frequency fluctuation correlation function, Eq. (\ref{eq:ccal}), and IR spectrum, Eq. (\ref{eq:irsp}).
(a): impact of different choices for initial time,
(b): impact of different choices for the end of the time interval,
(c): coupling strength between the NH vibration and the oscillator modes    as a function of their frequency (points are fitted results and line
    is guide for the eye),
(d): IR lineshape of NH stretching compared with experimental data \cite{woutersen04:5381} (line for experimental 
    result is drawn as guide for the eye only).
}
\label{fig:spec}
\end{figure}
\clearpage\newpage
\begin{figure}[ht]
\centering
\includegraphics[angle=-90,width=\textwidth]{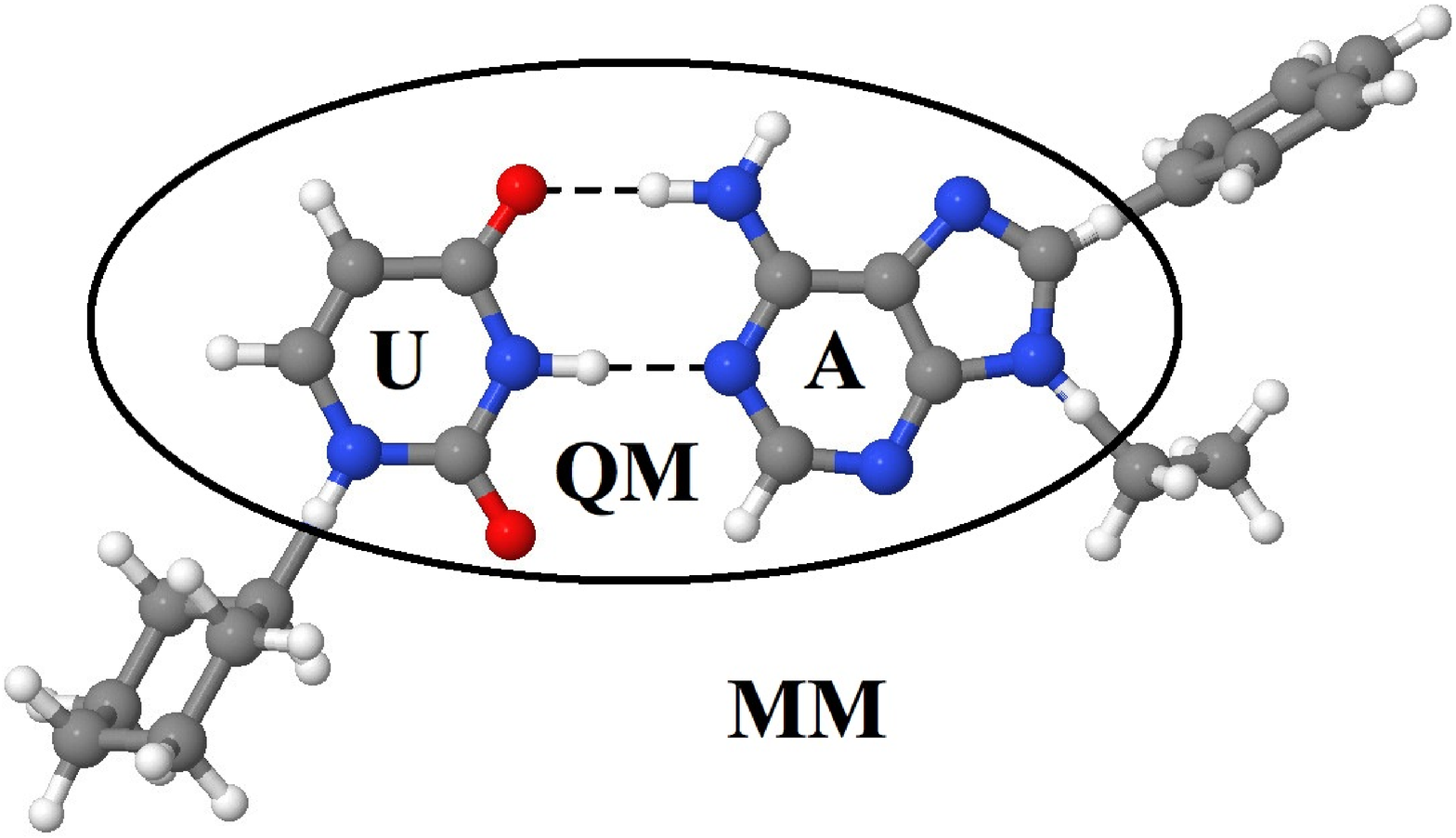} \\[15.0ex]
Fig. 1 (Yan et al.)
\end{figure}
\newpage
\begin{figure}[ht]
\centering
\includegraphics[width=0.9\textwidth]{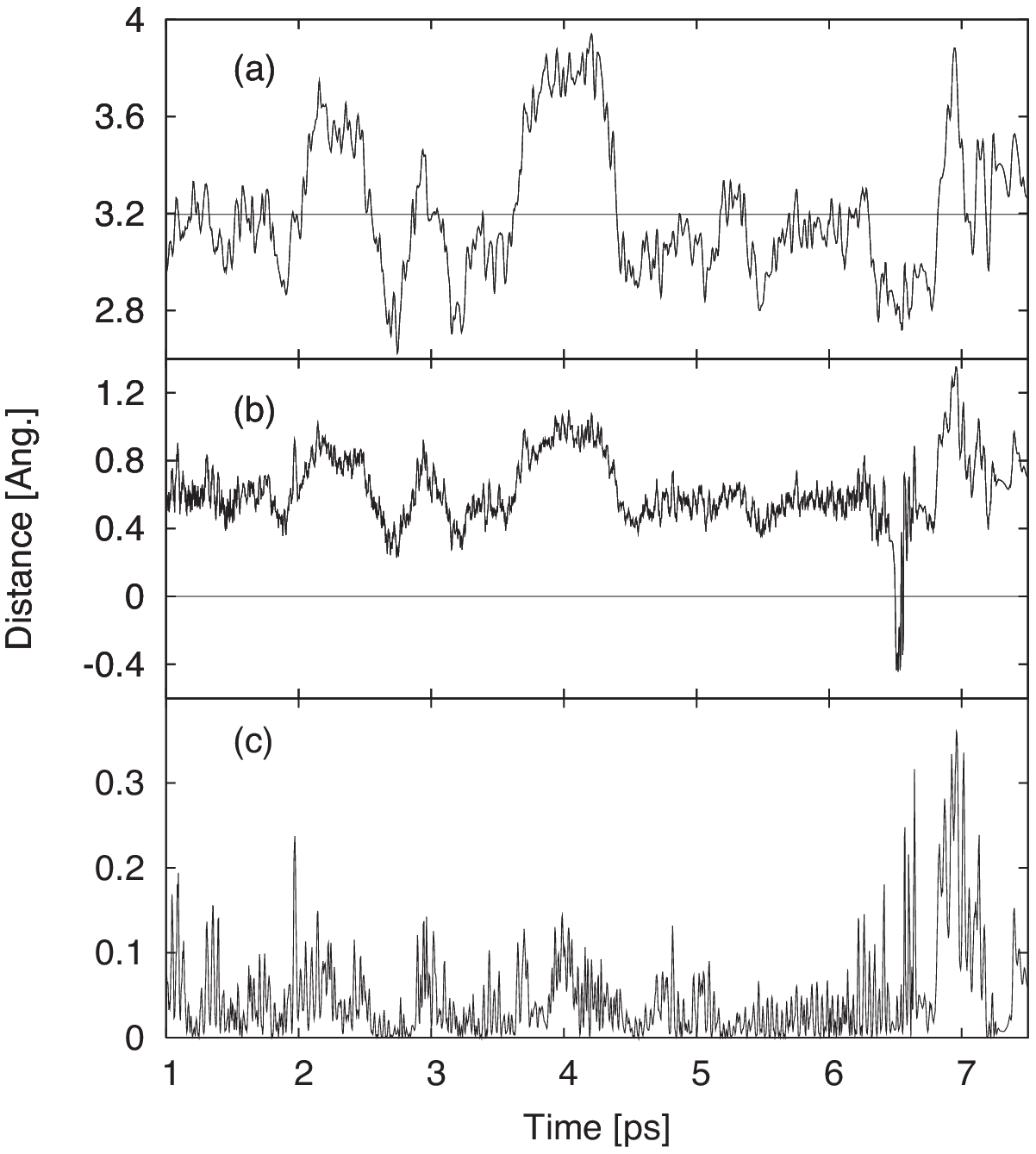}\\[15.0ex]
Fig. 2 (Yan et al.)
\end{figure}
\newpage
\begin{figure}[ht]
\centering
\includegraphics[width=0.9\textwidth]{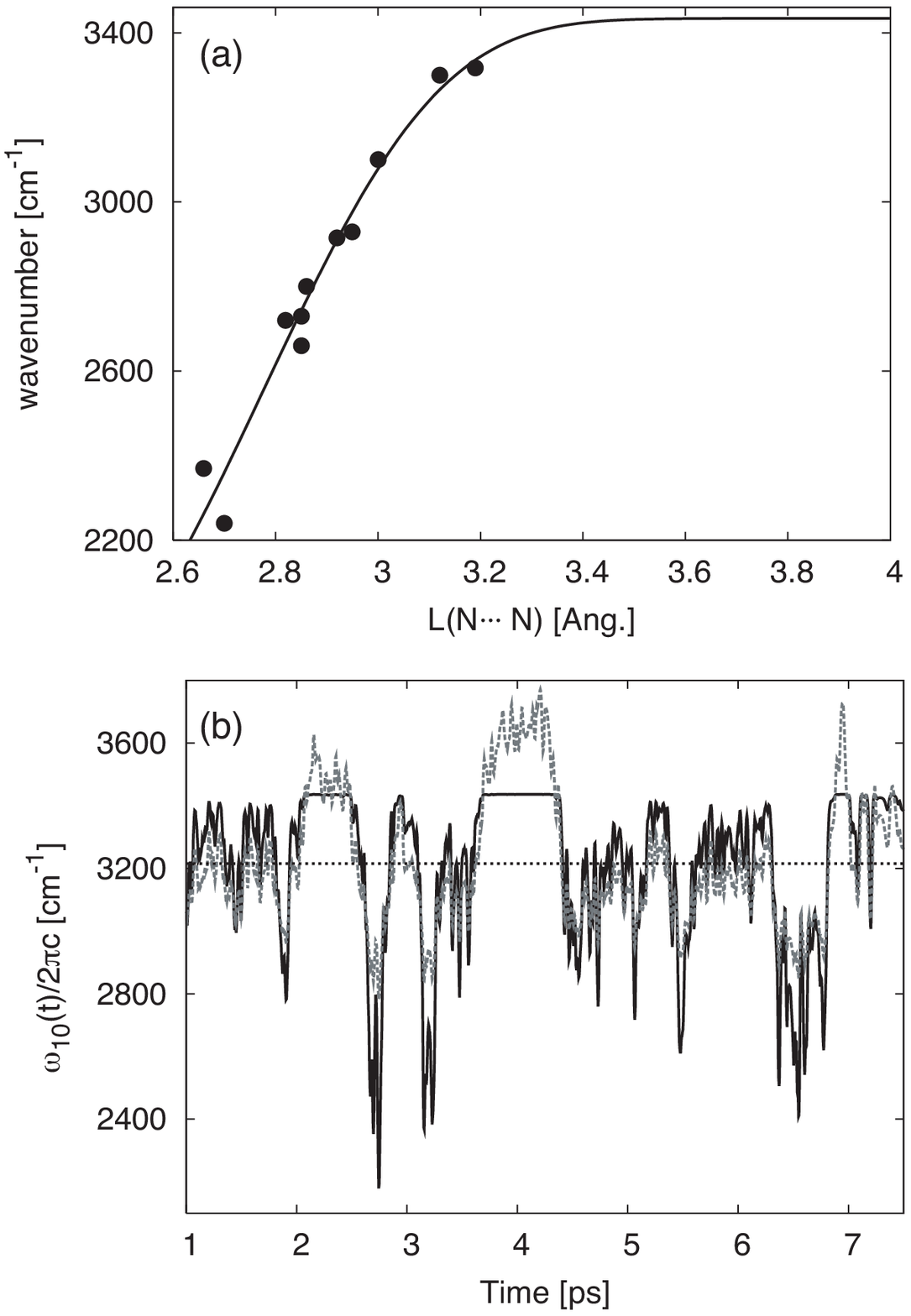}\\[15.0ex]
Fig. 3 (Yan et al.)
\end{figure}
\newpage

\begin{figure}[ht]
\centering
\includegraphics[width=\textwidth]{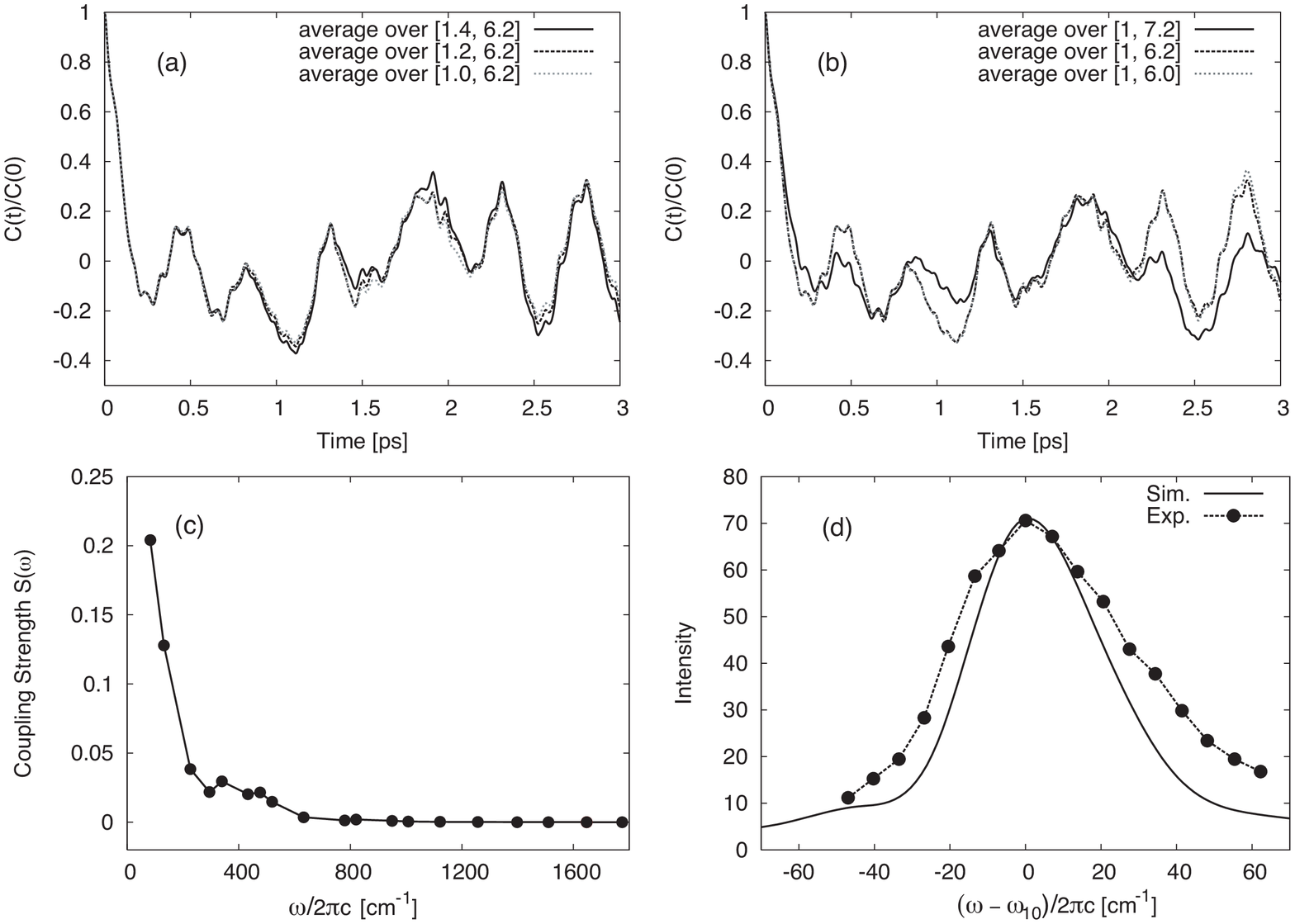} \\[15.0ex]
Fig. 4 (Yan et al.)
\end{figure}
\clearpage \newpage

%
\end{document}